\documentclass[fleqn]{article}
\usepackage{gc,amsfonts}
\def\mn{_{\mu\nu}}

\def\cK{{\cal K}}

\def\Z{{\mathbb Z}}

\def\GR{general relativity}

\def\GR{general relativity}

\def\bw{brane world}

\def\oV{\overline{V}}

\heads  {K.A. Bronnikov and B.E. Meierovich}
	{A general thick brane supported by a scalar field}

\begin{document}
\twocolumn[
\prepno{gr-qc/0402030}{\GC {9} 4, 313-318 (2003)}

\bigskip

\Title {A general thick brane supported by a scalar field}

\Authors{Kirill A. Bronnikov\foom 1} {and Boris E. Meierovich\foom 2}
{Center for Gravitation and Fundam.
    Metrology, VNIIMS, 3-1 M. Ulyanovoy St., Moscow 117313, Russia;\\
    Institute of Gravitation and Cosmology, PFUR,
        6 Miklukho-Maklaya St., Moscow 117198, Russia}
{Kapitza Institute for Physical
    Problems, 2 Kosygina St., Moscow 117334, Russia}

\Abstract
   {A thick $\Z_2$-symmetric domain wall supported by a scalar field with an
    arbitrary potential $V(\phi)$ in 5D general relativity is considered as
    a candidate \bw. We show that, under the global regularity requirement,
    such a configuration (i) has always an AdS asymptotic far from the
    brane, (ii) is only possible if $V(\phi)$ has an alternating sign and
    (ii) $V(\phi)$ satisfies a certain fine-tuning type equality. The thin
    brane limit is well defined and conforms to the Randall-Sundrum (RS2)
    \bw\ model if the asymptotic value of $V(\phi)$ (related to $\Lambda$,
    the effective cosmological constant) is kept thickness-independent.
    Universality of such a transition is demonstrated using as an example
    exact solutions for stepwise potentials of different shapes. One more
    result is that, due to scale invariance of the Einstein-scalar
    equations, any given regular solution creates a one-parameter family of
    solutions with different potentials. In such families, a thin brane
    limit does not exist while the ratio $\Lambda/\sigma^2$ ($\sigma$ is the
    brane tension) is thickness-independent and is in general different from
    its value in the RS2 model.
    }

]  
\email 1 {kb@rgs.mccme.ru}
\email 2 {meierovich@yahoo.com;\\ \hspace*{8mm}
	  http://geocities.com/meierovich/}

\section{Introduction}

   The \bw\ concept, suggested in the 80s \cite{ARVSG}, is broadly discussed
   nowadays in connection with the recent developments in
   superstring/M-theories \cite{Horava-Witten}. According to this concept,
   the standard-model particles are confined on a hypersurface, called a
   brane, which is embedded in a higher-dimensional space called the bulk.
   Various aspects of brane-world particle physics, gravity and cosmology are
   discussed in the recent review articles (\cite{Reviews}, see also
   references therein).

   Most of the studies are restricted to infinitely thin branes with
   delta-like localization of matter. This kind of models can, however, be
   only treated as an approximation since any fundamental underlying theory,
   be it quantum gravity or string theory, must contain a fundamental length
   beyond which a classical space-time description is impossible. It is
   therefore necessary to justify the infinitely thin brane approximation as
   a well-defined limit of a smooth structure, a thick brane, obtainable as
   a solution to coupled gravitational and matter field equations.

   In the context of the Universe evolution as a sequence of phase
   transitions with spontaneous symmetry breaking, a brane can be thought of
   as a plane topological defect. A systematic exposition of the potential
   role of topological defects in our Universe is provided by Vilenkin and
   Shellard \cite {Vilenkin and Shellard}. Particular types of defects:
   strings \cite {Strings}, monopoles \cite{Monopoles}, or domain walls are
   determined by the topological properties of vacuum \cite{Kibble}. These
   properties are well macroscopically described by using a scalar field
   with a proper symmetry-breaking potential as an order parameter.

   So, like many other authors, we try to describe a thick brane in the
   framework of 5D \GR\ as a domain wall separating two different states
   of a scalar field. We study analytically scalar field structures with
   arbitrary potentials, assuming $\Z_2$ symmetry with respect to the middle
   plane of the wall and restrict ourselves to Poincar\'e branes, i.e.,
   flat domain walls. Most of the existing problems show up quite clearly in
   these comparatively simple systems; moreover, in the majority of physical
   situations, the inner curvature of the brane is much smaller than the
   curvature related to brane formation, therefore the main qualitative
   features of Poincar\'e branes should survive in curved branes.

   Our work confirms and generalizes the well-known results obtained in a
   number of specific models [8--13]: 
   in the framework of 5D \GR, a globally regular thick brane always has an
   anti-de Sitter (AdS) asymptotic and is only possible if $V(\phi)$ has an
   alternating sign. Furthermore, if far in the bulk the potential tends to
   a fixed value independent of the brane thickness, then the thin brane
   limit is well defined and conforms to the celebrated RS2 \bw\ model
   \cite{RS2}, We demonstrate this with the aid of the simplest exactly
   solvable example: a potential consisting of two constant steps [see \eq
   (\ref{2-step})]. The RS2 limit, with the corresponding (fine-tuned)
   relation between the brane tension $\sigma$ and the bulk cosmological
   constant $\Lambda$, is found to be model-independent: it is obtained for
   arbitrary values of two shape factors of the potential (the steps'
   relative height and thickness) which are kept constant in the transition
   to zero thickness.

   However, not all sequences of brane-type solutions reproduce the RS2 thin
   brane limit. Our new observation is a scale invariance of the
   Einstein-scalar equations describing a thick brane. It puts into
   correspondence to any given solution a one-parameter family of solutions
   with different potentials. In any such family, the ratio
   $\Lambda/\sigma^2$ is fixed (thickness-independent) but a thin brane
   limit is meaningless since it leads to an infinite potential in the whole
   space.

   Quite evidently, to be considered as a model of our Universe, a brane
   like those discussed here must satisfy two major requirements: (i)
   ordinary matter should be confined to the brane to account for the fact
   that extra dimensions are not observed, and (ii) Newton's law of gravity
   should be reproduced on the brane in a non-relativistic limit. These
   issues, which have already been treated in a number of papers
   (among others, \cite{Reviews,b1,soda,b2}) 
   turn out to be quite nontrivial, and we hope to discuss them in the near
   future.

\section{Field equations and boundary conditions}

   We consider \GR\ in a 5D space-time, where we distinguish the usual
   four coordinates $x^\mu$, $\mu = 0,1,2,3$ and the fifth coordinate
   $x^4 = z$, to be used for describing the direction across the brane.
   The action is taken in the form $S = \int \sqrt{g} L\,d^5 x $, where
   $g = |\det (g_{AB})|$, $A,B = 0,1,2,3,4$ and $g_{AB}$ is the 5D
   metric tensor; the Lagrangian density has the form
\beq
    L = \frac{R}{2\kappa^2} + L_\phi, \cm
               L_\phi = \Half \d^A \phi\, \d_A\phi - V(\phi),    \label{L}
\eeq
   ($R$ is the 5D Ricci scalar and $\kappa$ the 5D gravitational constant)
   and leads to the Einstein-scalar equations
\bearr                                                          \label{EE}
   R_{A}^{B} - \half\delta_{A}^{B} R = -\kappa^2 T_{A}^{B}
                    = -\kappa^2 [\d_A\phi \d^B\phi - \delta_A^B\, L_\phi],
\yyy                                                          \label{eq_phi}
      g^{-1/2} \d_A (g^{1/2} g^{AB}\d_B \phi) = -d V/d\phi
\ear
   where \eq (\ref{eq_phi}) is a consequence of (\ref {EE}) due to the
   Bianchi identities.

   We seek regular solutions describing a static domain wall (thick brane),
   possessing $\Z_2$ symmetry with respect to the hypersurface $z=0$.

\subsection{Domain walls in flat space-time}

   Let us first consider, for comparison, the properties of a domain wall in
   flat 5D space-time with the metric $ds_5^2 = \eta\mn\,dx^\mu dx^\nu-dz^2$,
   $\eta\mn = \diag(1,\ -1,\ -1,\ -1)$ being the 4D Minkowski metric. The
   field equation (\ref{eq_phi}) for $\phi = \phi(z)$ reduces to
\beq                                                         \label{eq-flat}
    \phi'' = V_\phi \equiv dV/d\phi
\eeq
   where the prime stands for $d/dz$. As in many similar problems (say, for
   domain boundaries in ferromagnets), we assume the boundary conditions
\beq                                                         \label{Bound1}
    \phi (0) =0,  \cm  \phi(\pm \infty ) = \pm \phi_\infty
\eeq
    where $\phi_{\infty} = \const$. Then from (\ref{eq-flat}) it follows
    that $d V/d \phi =0$ at $\phi =\phi_{\infty}$. \eq(\ref{eq-flat}) has
    the first integral
\beq                                                  \label{1int-flat}
            \half (\phi')^2  = V(\phi) - V(\phi_{\infty})
\eeq
    and can be completely integrated:
\beq                                                 \label{z(fi)}
    z = \int_{0}^{\phi }\frac{d\phi}{\sqrt{2\left[ V(\phi) -V( \phi_{\infty})
                \right] }}.
\eeq

    Thus, for the Mexican hat potential
\beq                                                       \label{hat}
    V(\phi)=\fract{1}{4}\lambda(\phi^2 -\phi_{\infty}^2 )^2
\eeq
    the solution is written explicitly as
\beq                                                   \label{hat-solflat}
    \phi (z) =\phi_{\infty} \tanh
            \left( \sqrt{\lambda /2}\phi_{\infty} z\right) .
\eeq
    Another example is the periodic potential
\beq                                                      \label{V-cos}
        V(\phi) = \lambda \phi_{\infty}^{4}\cos ( \pi\phi/\phi_\infty),
\eeq
   for which the solution is
\beq                                                     \label{cos-sol}
    \phi (z) =\phi_{\infty}\biggl( 1-\frac{4}{\pi }\arctan
                \e^{-\pi \sqrt{\lambda }\phi_{\infty}z}\biggr).
\eeq

\subsection {Self-gravitating domain walls}

    Consider \eqs (\ref{EE}), (\ref{eq_phi}) for $\phi = \phi (z)$
    and the 5D metric
\beq                                                        \label{ds_gen}
     ds_5^2 = \e^{2F(z)} \eta\mn dx^\mu dx^\nu - \e^{8F(z)} dz^2,
\eeq
    where we have chosen the harmonic coordinate $z$, such that $\sqrt{g}
    g^{zz}=-1$. As a result, the 5D Ricci tensor acquires an especially
    simple form:
\bear                                                    \label{R_AB-harm}
     R_0^0 \eql R_1^1 = R_2^2 = R_3^3 = -\e^{-8F} F'',
\nn
     R_z^z \eql -4 \e^{-8F} (F'' - 3{F'}^2)
\ear
    The Kretschmann scalar ${\cal K} = R_{ABCD} R^{ABCD}$ is
\beq
      {\cal K} = 4\bigl[ \e^{-5F}(\e^{-3F} F')'\bigr]^2         \label{Kr}
                + 6\bigl( \e^{-4F} F'\bigr)^4.
\eeq
    For the metric (\ref{ds_gen}), $\cK$ is a sum of squares of all nonzero
    components $R_{AB}{}^{CD}$ of the Riemann tensor, therefore its finite
    value is necessary and sufficient for finiteness of all algebraic
    curvature invariants.

    The 5D Einstein equations (\ref{EE}) in our case reduce to
\bearr                                                   \label{F''=}
      F'' = -\frac{2\kappa^2 }{3}\e^{8F}V,
\yyy
      3(-F'' + 4 {F'}^2) = \kappa^2{\phi'}^2,           \label{no_V}
\yyy                                                \label{1int}
      {F'}^2 = \frac{\kappa^2 }{6}\left( \Half{\phi'}^2 -\e^{8F}V\right),
\ear
    where (\ref{1int}) is a first integral of the other two equations.
    The scalar field equation (\ref{eq_phi}) reads
\beq                                                    \label{phi''=}
            \phi'' = \e^{8F} d V/d\phi
\eeq
    and is also a consequence of (\ref{F''=}) and (\ref{no_V}). \eqs
    (\ref{F''=}) and (\ref{1int}) may be taken as a complete set of
    equations for $F(z)$ and $\phi (z)$; it is third-order and requires
    three boundary conditions.

    Now, the $\Z_2$ symmetry assumption dictates the boundary condition
    $F'(0) =0$. Then, assigning $F(0) =0$ by a proper choice of
    the time scale, we arrive at an unambiguous value of ${\phi'}^2$:
    ${\phi'}^2 (0) = 2V(0)$. So $F(z)$ is an even function while $\phi (z)$
    is an odd one. A complete set of boundary conditions
    at $z=0$, compatible with $\Z_2$ symmetry, is
\beq                                                      \label{Bound2}
    F(0) = F'(0) = 0,  \qquad  \phi (0) = 0.
\eeq

    Without gravity ($F=0$), \eq (\ref{phi''=}) reduces to (\ref{eq-flat}).
    However, in the gravitational case, for any fixed function $V(\phi)$,
    there is no free parameter in the equations and boundary conditions.
    It means that we now have no freedom to require that $\phi$ should tend
    to a constant value at $z\to \pm \infty$. Such a requirement is an
    additional constraint on the function $V(\phi)$, leading to a ``fine
    tuning'' between the brane and bulk parameters \cite{Shiromizu00}.

\section {Regularity conditions}

    Let us analyze the properties of solutions to \eqs
    (\ref{F''=})--(\ref{phi''=}) far from the brane, requiring a regular
    asymptotic at $z \to \infty$.

    Regularity of the metric [see (\ref{Kr})] implies $|F'|\e^{-4F} <
    \infty$, or $|b'(z)| < \infty$ where $b(z) \equiv \e^{-4F}$. The same
    space-time regularity requirement is translated to the scalar field SET
    via the Einstein equations, hence we should have everywhere
\beq                                                            \label{reg}
    |b'(z)| <\infty, \quad\  |V(\phi)| < \infty, \quad\
        				b(z)|\phi'(z)| < \infty.
\eeq

    \eq (\ref{no_V}) leads to $b''(z) > 0$. Since, according to our
    boundary conditions, $b'(0) = 0$, this means that $b(u)$ is an
    increasing function at $z > 0$, inevitably growing to infinity at
    large $z$ at least linearly for any solution with $\phi \ne \const$.
    The growth is precisely asymptotically linear due to (\ref{reg}),
    $b' \to \const >0$, and hence $F' \approx  -1/(4z)$ at large $z$.

    Returning to (\ref{F''=}) and integrating, we obtain
\beq                                                     \label{F'-as}
    F'(\infty) = -\frac{2}{3}\kappa^2
                              \int_{0}^{\infty}\e^{8F} V\,dz.
\eeq
    For regular solutions
\bearr                                                   \label{oV}
    \oV (\infty) =0, 
\nnn
     \oV (z) := \int_{0}^{z}\!\sqrt{g} V(z_1)\, dz_1 =
                              \int_{0}^{z}\!\e^{8F(z_1)} V(z_1)\,dz_1.
\ear
    This is the above-mentioned fine-tuning condition in terms of the
    potential $V$. The integral $\oV(z)$ is the invariant full potential
    energy per unit 3-volume in the layer from zero to $z$. Since
    $\e^{8F} = 1/b^2 > 0$, a nontrivial potential $V(\phi)$ must change its
    sign at least once to yield $\oV (\infty) = 0$.

    It is easy to show that in regular solutions $\phi' = o(1/z)$.
    Indeed, as follows from (\ref{reg}), $|\phi'|$ behaves at most as $1/z$,
    but in this case \eq (\ref{no_V}) leads to $b'' \sim 1/z$, hence (taking
    into account that $b'' > 0$) we would have $b'\to\infty$ contrary to
    (\ref{reg}).

    We conclude that $\e^{-4F}\phi' \to 0$ at large $z$, and \eq
    (\ref{F''=}) shows that $V$ tends to a finite negative value. If
\beq                                                           \label{F-as}
    b(z)=\e^{-4F} \approx kz, \quad\ k = \const > 0, \quad\ z\to \infty,
\eeq
    then
\beq
    \kappa^2 V\Big|_{z=\infty} = \Lambda =- \frac{3k^2}{8},    \label{V-as}
\eeq
    where $\Lambda$ is the effective cosmological constant.

    Consider the scalar field behaviour in more detail.
    Since $\phi' = o(1/z)$, it is reasonable to suppose that $\phi$ tends to
    a finite value $\phi_\infty$. Then a small deviation
    $\phi_1 = \phi -\phi_\infty$ obeys the linear equation
\beq                                                      \label{fi1''=}
     \phi''_1 = \e^{8F} V_{\phi\phi} (\phi_\infty)\cdot \phi_1.
\eeq
    We took into account that $V_\phi (\phi_\infty) =0$, which follows from
    (\ref{phi''=}) due to $\phi'=o(1/z)$. Now, one can verify that \eq
    (\ref{fi1''=}) has solutions vanishing as $z \to \infty$ only in case
    $V_{\phi\phi} = m^2 > 0$, i.e., when $\phi_\infty$ is a minimum of
    $V(\phi)$, and $m$ is then an effective mass of the small linear field
    $\phi_1$. Substituting $\e^{8F}$ from (\ref{F-as}) into (\ref{fi1''=}),
    we obtain at large $z$
\bearr                                                     \label{fi1''...}
    \phi_{1}'' = \frac{B}{z^2 }\phi_{1},
\cm
     B = \frac{3 m^2}{8\kappa^2 |V(\phi_\infty)|}= \frac{3m^2}{8|\Lambda|}.
\ear
    Its solution, vanishing as $z \to \infty $, is
\[
     \phi_{1}\sim \frac{1}{z^{p-1}},\cm  p=\frac{1}{2}
                           \left(1+\sqrt{1 + 4B}\right),
\]
    and, as we need, $\phi'\sim z^{-p} = o(1/z)$ since $p > 1$.

    We have obtained that the potential $V(\phi)$ changes its sign at least
    once and tends to a negative value as $z\to \infty$, where $V(\phi)$ has
    a minimum: $V_\phi =0$ and $V_{\phi\phi} > 0$, and the integral $\oV$
    (\ref{oV}) is zero.  Such a minimum can in principle occur at finite $z$,
    but a minimum for which $\oV(z) = 0$ (that is, at which $F'=0$) can take
    place only at $z=\infty$. In other words, one cannot assert that the
    minimum of $V$ corresponding to a regular asymptotic is a minimum
    nearest to $\phi=0$, but it is the nearest at which $\oV =0$.

    This behaviour is not unique: despite $\phi' = o(1/z)$, one cannot
    exclude that $\phi \to \infty$, though slower than $\ln z$ (it can be,
    for instance, $\phi' \sim 1/(z\ln z)$ and $\phi \sim \ln\ln z)$. Then
    $V (\phi)$ tends to the value (\ref{V-as}) as $\phi \to \infty$.

    In any case, due to oddness of $\phi (z)$, the values $\phi (+\infty) =
    -\phi(-\infty) \ne 0$ make the domain wall topologically stable. The
    postulated $\Z_2$ symmetry implies that $V(\phi)$ is an even function.
    Its asymptotic value, $V(z = \pm\infty) = \Lambda/\kappa^2 < 0$ plays
    the role of a cosmological constant at the bulk asymptotic, and the
    metric is asymptotically anti-de Sitter (AdS). The values $z = \pm\infty$
    then correspond to an anti-de Sitter horizon.

\section{Brane tension and the thin brane limit}

    Consider the thin brane limit of regular solutions with finite
    $\phi_\infty$, leaving aside the above case of slowly growing $\phi(z)$.
    We thus have $V(\phi)$ with a minimum at $\phi = \phi_\infty$, and there
    holds the ``fine tuning'' condition $\oV(\infty) =0$.

    The action $S$ with the Lagrangian $L = L_G + L_\phi$
    may be split into the bulk and brane parts,
\bear  \nq                                                     \label{S_tot}
    S \eql S_{\rm bulk} + S_{\rm brane},
\\     \nq                                                    \label{S_bulk}
    S_{\rm bulk} \eql -\int \frac{R - 2\Lambda}{2\kappa^2 }
     \sqrt{g}d^{5}x,\quad\  \frac{\Lambda}{\kappa^2} = V(\phi_\infty),
\\     \nq                                                   \label{S_brane}
    S_{\rm brane} \eql \int \!\left[ \frac{1}{2}
           \d_A\phi\, \d^A\phi + V(\phi_\infty) -V(\phi) \right]
               \! \sqrt{g}\,d^{5}x.
\nnn
\ear
    The brane action may be presented in the form
\bearr                                                       \label{tension}
     S_{\rm brane} = - \int \sigma \,d^4 x,
\nnn
    \sigma = \int \left[- \frac{1}{2}\d_A\phi\, \d^A\phi
                    + V(\phi) - V(\phi_\infty) \right] \e^{8F} dz,
\ear
    where the quantity $\sigma$ can be regarded as the brane tension. It is
    equal to the total scalar field energy per unit 3-volume on the brane,
    in which the potential energy is counted from the vacuunm level
    $V_\infty$.

    The second Randall--Sundrum (RS2) model of a thin brane \cite{RS2} is
    based on the splitting (\ref{S_tot}): assuming a delta-like matter
    distribution characterized by the tension $\sigma$ and using the
    Israel matching condition for the 5D metric, they found the
    fine-tuning condition
\beq                                                      \label{RS_fine}
     6 \Lambda =  -\kappa^4\sigma^2.
\eeq

    In our approach, a transition to a thin brane can be carried out along a
    sequence of solutions with different potentials but a fixed value of
    $\Lambda$, which characterizes the background properties of 5D
    space-time and determines a length scale $1/\sqrt{\Lambda}$ independent
    of the brane thickness. This corresponds to a brane as a domain wall
    between two vacua with equal and fixed energy densities.

    Then, if the thin brane concept is correct, we should
    expect that, independently of the specific form of the potential,
\beq                                                      \label{thin_lim}
    \lim\limits_{a \to 0} \frac{|\Lambda|}{\kappa^4 \sigma^2} = \frac 16,
\eeq
    the parameter $a$ characterizing the brane thickness.

    It is hard to prove the consistency of the thin brane limit for models
    with arbitrary potentials. Let us therefore consider the simplest
    potential admitting an exact solution, containing an explicit thickness
    parameter and simulating a variety of shapes of $V(\phi)$. Namely,
    introducing more convenient variables
\beq                                                       \label{def_fv}
       f(z) := \frac{2\kappa}{\sqrt{3}} \phi(z), \cm
       v(f) := \frac{8}{3} \kappa^2 V(\phi),
\eeq
    consider the following two-step potential:
\beq
      v = v(z) = \vars{ v_1 = \const > 0, &\quad 0 < z < c,\\ \label{2-step}
                    v_2 = \const > 0, &\quad c < z < a,\\
                v_3 = \const < 0, &\quad z > a.   }
\eeq
    It enables us to study the possible dependence of the solutions on the
    shape of the potential. The thin brane limit is reached when $a\to 0$,
    and in this transition one can preserve the character of the potential by
    keeping constant the ratios $c/a$ and $v_1/v_2$. Both $v_1 > 0$ and $v_2
    > 0$ are taken for simplicity, it is quite easy to include the
    possibility $v_1 \leq 0$ or $v_2 \leq 0$ [but at least one of them should
    be positive to satisfy (\ref{oV})].

    \eqs (\ref{F''=})---(\ref{1int}) are rewritten as
\bear
    v \eql bb'' - b'^2 \,=\, b^2 f'^2 - b'^2,                 \label{v(z)}
\\
    f'^2 \eql b''/b.                                     \label{f'(z)}
\ear
    The boundary conditions at $z = 0$ are
\beq
    b(0) = 1, \cm b'(0) = 0, \cm f(0) = 0.              \label{bound_0}
\eeq
    For the potential (\ref{2-step}), \eqs (\ref{v(z)}), (\ref{f'(z)}) have
    the following solution satisfying (\ref{bound_0}) and regular at large
    $z$:
\bear \nq                                                   \label{sol-ex2}
    z < c: \quad \lal b(z) = \cosh (k_1 z),
                         \quad\  f'(z) = k_1 = \sqrt{v_1};
\nn   \nq\,
    c < z < a: \quad \lal b(z) = \frac{\sqrt{v_2}}{k_2}\cosh [k_2(z{-}z_2)],
                \quad\  f'(z) = k_2;
\nn   \nq
    z > a: \quad  \lal  b(z) = \sqrt{-v_3} (z-z_3), \quad\ f(z) = \const,
\nnn
\ear
    where $k_1,\ k_2,\ z_2,\ z_3$ are integration constants, and the proper
    smoothness of the solution is achieved by matching the values of $b(z)$,
    $b'(z)$ and $f(z)$ at $z = a$ and $z = c$. This gives:
\bear
    \cosh (k_1 c) \eql \frac{\sqrt{v_2}}{k_2}\cosh [k_2(c-z_2)];\label{match}
\nn
    k_1 \tanh (k_1 c) \eql k_2 \tanh [k_2(c-z_2)];
\nn
    \frac{\sqrt{v_2}}{k_2}\cosh [k_2(a-z_2)] \eql \sqrt{-v_3} (a-z_3);
\nn
    k_2 \tanh [k_2(a-z_2)] \eql 1/(a-z_3).
\ear

    The fine-tuning condition (\ref{oV}) is satisfied automatically. The
    tension $\sigma$ is calculated as follows:
\bearr                                                      \label{sig-ex2}
     \frac{4}{3}\kappa^2 \sigma = \int_0^\infty dz
            \biggl[\frac{|v_3|}{b^2 (z)} + f'^2(z)\biggr]
\nnn \ \
        = k_1^2 c + k_2^2 (a-c) + \frac{1}{a-z_3}
                      + \frac{|v_3|}{k_1} \tanh (k_1 c)
\nnn \ \
        + \frac{|v_3|k_2}{v_2} \Bigl\{
             \tanh [k_2(a - z_2)] - \tanh [k_2(c - z_2)] \Bigr\}.
\ear

    Quite evidently, this expression essentially depends on all parameters of
    the model. The situation changes when we pass to the thin brane limit,
    $a\to 0$, keeping constant $v_3$, $v_1/v_2$ and $c/a$. In this case
    different quantities have the following orders of magnitude:
\bearr
        k_1 \sim k_2 \sim 1/\sqrt{a};    \cm\ \   v_1 \sim v_2 \sim 1/a;
\nnn
        a-z_2 \sim a, \cm \qquad\      a-z_3 = O(1).
\earn
    The matching conditions (\ref{match}) lead to
\bearr                                                      \label{match-0}
            k_2 \approx \sqrt{v_2};
     \cm        \frac{c-z_2}{c} \approx \frac{v_1}{v_2};
\nnn
        \frac{1}{a-z_3} \approx v_2 (a-z_2) \approx \sqrt{|v_3|}.
\ear
    Substituting all this to (\ref{sig-ex2}), we see that the last three
    terms with hyperbolic tangents vanish as $a\to 0$ while the remaining
    ones give
\beq                                                        \label{sig0-ex2}
     \frac{4}{3}\kappa^2 \sigma \approx
                2 [v_1 c + v_2 (a-c)] \approx 2 \sqrt{|v_3|}
\eeq
    Recalling that $v_3 = (8/3)\Lambda$, we see that the thin-brane relation
    (\ref{RS_fine}) is restored irrespective of the shape factors $a/c$
    and $v_1/v_2$. This testifies the universality of the thin brane limit of
    thick branes.

\section {Scale invariance}

\def\ta{\tilde{a}}
    The Einstein equations (\ref{F''=})--(\ref{1int}) or, equivalently,
    (\ref{v(z)}), (\ref{f'(z)}) are invariant under the scaling
    transformation
\beq
      z \to \ta z, \qquad\  v \to v/\ta^2 .              \label{s-trans}
\eeq
    Applying this transformation to any particular solution, one obtains
    from it a one-parameter family of solutions with different
    potentials\footnote
{Both $a$ from the previous section and $\ta$ are related to brane thickness.
 However, they parametrize quite different sequences of solutions. In the
 $a$-sequence the value of $\Lambda$ is kept constant whereas in the
 $\ta$-sequence we have $\Lambda \sim \ta^{-2}$.}.
    One can note that the transformation (\ref{s-trans}) looks so simply due
    to our choice of the harmonic coordinate $z$; with other coordinates,
    this scale invariance would take a less transparent form.

    As functions of $\ta$, $\Lambda \sim V( \phi_{\infty}) \sim \ta^{-2}$
    while the tension $\sigma \sim \ta^{-1}$, hence $\Lambda \sim \sigma^2$.
    However, the $\ta$-independent ratio $|\Lambda|/(\kappa^4\sigma^2)$
    (which is 1/6 in the RS2 case) is model-dependent, in other words, it
    depends on the shape of the potential and is, in general, unequal to 1/6.

    This important circumstance is not directly related to the thin brane
    limit of the models under study since there (see above) the transition
    is carried out along a sequence of solutions with a fixed value of
    $\Lambda$, which characterizes the background vacuum energy density and
    determines a length scale ($1/\sqrt{\Lambda}$) independent of the brane
    thickness.

    Consideration of a $\ta$-parametrized sequence of solutions is more
    appropriate if the brane is treated as the principal source of the
    geometry. The very existence of the bulk cosmological constant $\Lambda$
    is then caused by the brane. In this case the thin brane limit is
    meaningless ($\ta\to 0$ implies $\Lambda \to \infty$). At any rate, for
    a classical macroscopic description, the brane thickness should be
    greater than the Planck length.

    Let us give a specific example using the inverse problem method:
    given $b(z)$ (or $F(z)$), one easily finds $v(z)$ and $f(z)$ from
    \eqs (\ref{v(z)}) and (\ref{f'(z)}). According to (\ref{f'(z)}), one must
    choose such $b(z)$ that $b'' > 0$, then $f(z)$ will be monotonic, leading
    to a well-defined dependence $v(f)$, or $V(\phi)$.

    Consider the function
\beq                                                    \label{b-ex1}
        b(z) = (1 + z^2/\ta^2)^{1/2}.
\eeq
    It satisfies the boundary conditions (\ref{bound_0}) and leads to the
    asymptotic (\ref{F-as}). With (\ref{b-ex1}), the equations give
\beq                                                    \label{v-ex1}
    v(z)= - \frac{z^2 - \ta^2}{a^2 (z^2 + \ta^2)};
\qquad
    f(z) = \arctan \biggl(\frac{z}{\ta}\biggr).
\eeq
    Excluding $z$, we find that it is a cosine potential:
\beq                                                    \label{v=cos}
    v(f) = \ta^{-2} \cos (2f).
\eeq
    This solution was previously found by Gremm \cite{Gremm1}.
    The limiting value of $\phi$,
    $\phi_\infty = (\pi/4)\sqrt{3/\kappa^2}$,
    is unambiguously related to the 5D gravitational constant
    $\kappa^2$ (fine tuning), while the ``thickness parameter'' $\ta$
    remains arbitrary. The constant $\Lambda$ is
\beq                                                        \label{Lam}
     \kappa^2 V\Big|_{z=\infty} = \frac{3}{8}\,v\Big|_{z=\infty}
            = -\frac{3}{8\kappa^2 \ta^2}.
\eeq
    As was predicted, the relation between $\Lambda$ and $\sigma$
    is $a$-independent:
\[
    \frac{\Lambda}{\sigma^2} = -\frac{32\kappa^4}{27\pi^2}.
\]
    The model-dependent numerical factor $\frac{32}{27\pi^2} = 0.120...$
    is about 30 per cent smaller than the RS2 value 1/6.

    A thin brane limit of this solution for fixed $\Lambda$ cannot be found
    since for given $\Lambda$ it has no free parameter. A more complex
    family of solutions including the present one as a special case was
    described by Melfo et al. \cite{melfo}, who also asserted that it was a
    regularized version of the RS2 model which could be restored by
    turning their thickness parameter to zero.

\section {Concluding remarks}

    We have performed a general study of regular domain walls (thick branes)
    supported by a minimally coupled scalar field with an arbitrary
    potential in 5D \GR. It has been shown that the {\it only\/} kind of
    asymptotic for such walls is AdS and it is only possible if $V(\phi)$ has
    an alternating sign and satisfies the fine tuning condition (\ref{oV}).
    We have confirmed that the zero thickness limit of such branes is well
    defined and conforms to the RS2 \bw\ model. This follows from the
    properties of an exact solution for the stepwise potential (\ref{2-step})
    which contains two free shape factors $c/a$ and $v_1/v_2$ kept constant
    as the thickness $a\to 0$. Other explicit examples of such a transition
    with some special potentials $V(\phi)$ were studied previously
    (see \cite{melfo,ghoroku} and references therein) with the same result.
    Although our set of stepwise potentials is far from being general, it
    encompasses quite various shapes of $V(\phi)$; moreover, it can be
    extended to any number of steps and, in the limit of small steps, can
    approximate any continuous potential and even many discontinuous ones.
    The existence of a correct thin brane limit can probably be proved for a
    very broad class of scalar field potentials in the same manner as one
    proves the existence of a definite integral in the calculus.

    We have also shown that in a regular, asymptotically AdS configuration
    the scalar field $\phi$ not necessarily tends to a constant value at
    large $z$ and can very slowly [as $o(\ln z)$] but infinitely grow at
    large $z$. In such models, $V(\phi) \to \Lambda =\const <0$ as $\phi\to
    \infty$.

    The present stringent asymptotic conditions for regular solutions are
    not necessary if there are at least two branes in the bulk, as is the
    case in the first Randall-Sundrum (RS1) model \cite{RS1} where the fifth
    dimension is compact and there is no asymptotic at all. The asymptotic
    conditions can also change if we cancel the $\Z_2$ symmetry condition,
    as is shown, in particular, in Ref.\,\cite{melfo}.

\Acknow{KB acknowledges partial financial support from ISTC Proj. \#\ 1655.}

\end{document}